\documentclass[runningheads]{llncs}
\AtBeginDocument{%
  \providecommand\BibTeX{{%
    \normalfont B\kern-0.5em{\scshape i\kern-0.25em b}\kern-0.8em\TeX}}}

\usepackage{url}
\usepackage{graphicx}

\newcommand{\partitle}[1]{\vspace{2mm}\noindent\textbf{#1}}
\usepackage{subfig}
\begin{document}

\title{The Impact of User Demographics and Task Types on Cross-App Mobile Search}
\titlerunning{Impact on Cross-App Mobile Search}

\author{Mohammad Aliannejadi\inst{1}\orcidID{0000-0001-5134-5234}
\and
Fabio Crestani\inst{2}\orcidID{0000-0001-8672-0700}
\and
Theo Huibers\inst{3}\orcidID{0000-0002-9837-8639}
\and
Monica Landoni\inst{2}\orcidID{0000-0003-1414-6329}
\and
Emiliana Murgia\inst{4}
\and
Maria Soledad Pera\inst{5}\orcidID{0000-0002-2008-9204}}

\authorrunning{M.~Aliannejadi et al.}
%
\institute{University of Amsterdam, Amsterdam, The Netherlands \\ \email{m.aliannejadi@uva.nl}
\and
Universit\`a della Svizzera Italiana, Lugano, Switzerland \\
\email{fabio.crestani@usi.ch, monica.landoni@usi.ch}
\and
University of Twente, Enschede, The Netherlands \\
\email{t.w.c.huibers@utwente.nl}
\and
Universit\`a degli Studi di Milano-Bicocca Milano, Italy \\
\email{emiliana.murgia@unimib.it}
\and
Boise State University, Boise, Idaho, USA \\
\email{solepera@boisestate.edu}
}


\maketitle

\begin{abstract}

Recent developments in the mobile app industry have resulted in various types of mobile apps, each targeting a different need and a specific audience. Consequently, users access distinct apps to complete their information need tasks. This leads to the use of various apps not only separately, but also collaboratively in the same session to achieve a single goal. Recent work has argued the need for a \emph{unified mobile search} system that would act as metasearch on users' mobile devices. The system would identify the target apps for the user's query, submit the query to the apps, and present the results to the user in a unified way. In this work, we aim to deepen our understanding of user behavior while accessing information on their mobile phones by conducting an extensive analysis of various aspects related to the search process. In particular, we study the effect of task type and user demographics on their behavior in interacting with mobile apps. Our findings reveal trends and patterns that can inform the design of a more effective mobile information access environment.


\end{abstract}

\section{Introduction}
The ever-increasing use of smartphones has made them pervasive in our lives, originating an abundance of mobile apps that users install and use~\cite{Crestani2017}. Many of the apps that users interact with daily have their own data repository and feature their own search engine. This prompted researchers to study and report on the need and significance of having a truly universal mobile search framework that would act as a metasearch engine on the device~\cite{DBLP:conf/sigir/AliannejadiZCC18,DBLP:conf/cikm/AliannejadiZCC18,DBLP:journals/corr/abs-2101-03394}. In this case, users could type their search queries in a unique search box and the framework would route the query to relevant apps that could retrieve useful results that would then be displayed in a unified interface. To inform the design of such an engine, it is critical to understand how users interact and access information using different apps. It is also imperative to understand user behavior while accessing different apps on their smartphones, as this plays a crucial role in improving the system. 


The high significance of understanding user behavior in relation to various demographic attributes and the prominence of cross-app search in people's lives, motivated us to study how different users interact with different apps as they complete a search task. While the influence of user demographics on web search queries~\cite{DBLP:conf/sigir/WeberC10} and app usage \cite{DBLP:conf/icwsm/MalmiW16,DBLP:journals/percom/ZhongTM013} has been already investigated, to the best of our knowledge no work has looked at cross-app search queries. In this paper, we study the behavior of over 600 users in terms of mobile app usage  over 200 search tasks to answer two research questions: {\em RQ1}: Do demographic factors condition app usage for search? and, {\em RQ2}: Do extrinsic factors impact app usage for search? We analyze the relationship between users' app selection behavior with respect to different demographics characteristics as well as other system-related  aspects. In  particular, we study age, education, device type, and task type. We observe that all of these dimensions impact the way users complete search tasks on their smartphones.

Trends and patterns emerging from the analysis we conducted reveal the impact that demographic factors have on users' selection to conduct information-seeking tasks; the device and the task type itself also direct app selection. Findings from this work could serve as groundwork informing the design of a personalized metasearch system for mobile devices; they also offer insights that recommender systems could leverage in terms of suggesting suitable apps given the task of choice, in addition to diversification of app selection to complete search tasks on mobile devices.  







\section{Related Work}
In this section, we discuss existing literature that offers context to our work. We first emphasize the influence that demographic information has on several areas of study. We then briefly mention existing works focused on mobile app search.

Existing works have emphasized the importance of demographic information from various perspectives~\cite{hinds2018demographic} such as web search~\cite{DBLP:conf/sigir/WeberC10}, video consumption~\cite{DBLP:conf/emnlp/WangXMX16}, music~\cite{DBLP:journals/mta/KrismayerSKR19}, and mobile app usage~\cite{DBLP:conf/icwsm/MalmiW16,DBLP:journals/percom/ZhongTM013}. These studies reveal that understanding users' demographics and usage patterns is crucial to provide enhanced service and identify which users to target (e.g., showing ads about family vacations). Weber and Castillo \cite{DBLP:conf/sigir/WeberC10} studied the behavior of different user segments on web search from various aspects (e.g., income level, education, ethnicity) looking at how demographic aspects affected their search queries and clicks. They demonstrated the high impact that demographics can have on user behavior. This motivated a series of works aiming to predict user demographics based on user behavior, and in this context, mobile app usage has been extensively studied~\cite{DBLP:conf/icwsm/MalmiW16,DBLP:journals/percom/ZhongTM013}. 

More closely focusing on mobile search, we start with the study in \cite{ong2017using}, which outlines differences on search behavior observed on mobiles vs desktop Web search. 
In a similar work, Song et al.~\cite{DBLP:conf/www/SongMWW13} found a significant difference in search patterns done using iPhone, iPad, and desktop.
We also highlight the work by Carrascal and Church \cite{carrascal2015situ}, who examine users' engagement with mobile search and report that in a mobile context, users turn to more apps and that certain app categories are used more intensively. 
Kamvar et al.~\cite{DBLP:conf/chi/KamvarB06} did a large-scale mobile search query analysis, finding mobile search topics were less diverse. Similar studies done in \cite{DBLP:conf/sigir/Guy16} and \cite{DBLP:journals/jasis/CrestaniD06} compared typed-in queries and spoken queries on mobile devices conducted comparative studies on mobile spoken and typed-in queries where they found similar conclusions, i.e., spoken queries are more similar to natural language.
Tian et al. \cite{tian2020identifying} look into how automatic task segmentation can directly improve mobile search. For instance, the authors discuss how after performing a certain task the probability to formulate a particular query increases \cite{tian2020identifying,zhang2016towards}.

This work is closely related to our previous studies on unified mobile search~\cite{DBLP:conf/sigir/AliannejadiZCC18,DBLP:conf/cikm/AliannejadiZCC18,DBLP:journals/corr/abs-2101-03394} where we introduced and studied research on unified mobile search and collected cross-app queries through crowdsourcing, as well as in situ user study. We based our work here on the data collected through an in situ user study and provide further analysis on how users with different demographics interact with applications while searching on mobile devices.

\section{Methodology}
In this section, we describe the data and experiments that we used to address our research questions.

\subsection{Data}
We use the UniMobile\footnote{\url{https://github.com/aliannejadi/unimobile}} dataset released by Aliannejadi et al. \cite{DBLP:conf/sigir/AliannejadiZCC18,DBLP:conf/chiir/CostaAC20}. The dataset contains 5,812 cross-app mobile search queries for 206 search tasks spread across multiple task categories. The dataset also includes demographics surveys in which participants provided details about their background, search experience, and preferences. They also answered survey questions aimed at understanding how participants access the Internet and use their phones. In particular, in one question the participants specified the device that they most frequently used to access the Internet. Finally, participants shared whether they use their smartphones primarily for personal reasons or work-related reasons.

The data contains the queries submitted by 625 users located in the United States (400 identified themselves as female, the rest as male). 39\% of the participants were aged between 25 to 34, followed by 24\% between 35 and 44, and 20\% between 18 and 24. 17\% of the participants were in other age groups. Most participants held a Bachelor's degree (38\%), followed by ``Some college, no degree'' (26\%), and Master's degree (16\%). The other 20\% had other levels of education (e.g., high school and doctorate). 

The majority of participants used their smartphones as their primary Internet device (48\%), followed by 31\% using a Laptop computer, 14\% Desktop computer, 5\% tablets, and 2\% other devices. 
Most participants stated that they used their smartphones more for personal reasons (71\%), 19\% about an equal amount for work and personal reasons, and 10\% more for work-related reasons.

\subsection{Experiments}
\label{sub:expSetUp}
Here, we describe the experiments we designed based on user demographics and task-related measures.

\partitle{Age.}
To investigate the effect of users' age on their search behavior, in Figure~\ref{fig:age_count} we show the distribution of selected apps per age group. We count the total number of unique apps that each participant selects when completing different tasks and plot the distribution of each age group. We hypothesize that users of different age groups use a different range of apps to complete their daily search tasks.

\partitle{Education.}
We examine the effect of education from three different perspectives. First, in Figure~\ref{fig:edu_count} we show the distribution of the total number of unique apps per user. Our hypothesis is that users' educational background plays a role in their app selection. Second, in Figure~\ref{fig:diversity_edu} we show the diversity of users selecting apps for different tasks. For each user, we count the number of times they select each app for different tasks. Then, we plot the unique app count per user group (and 95\% confidence interval)\footnote{Estimated via empirical analysis of a Bootstrap sampling with 1,000 resamples.}. We hypothesize that users' ability and/or tendency to diversify their app selection depends on their level of education. Third, we are interested in finding out if education impacts how much users would ``follow the crowd.'' In other words, we observe whether a user selects an app that the majority of users have also selected for the same task (e.g., if 8 out of 10 users choose Google Search for a task, would a user choose the same app or not?). We show in Figure~\ref{fig:edu_25} the number of apps that are not in line with the crowd per user (and 95\% confidence interval). To determine popular apps for a given task, we first assume that the number of app selections follows a normal distribution. Then, we consider apps that fall into the 25\textsuperscript{th} quantile as \textit{rare}, i.e., non-popular. If a user chooses one of the popular apps for a given task, we treat this as a \textit{follow-the-crowd} selection and dismiss it in our counts. Instead, if a user selects one of the  ``rare" apps, we include that in our computation.  

\partitle{Device.}
We plot in Figure \ref{fig:count_device} app usage across devices used to access the Internet, whereas in Figure~\ref{fig:diversity_device} we capture how many apps users choose to complete different tasks with (and 95\% confidence interval). Our hypothesis is that the preferred device to access the Internet influences the user's behavior.

\partitle{Task.}
We study task impact in two different experiments. First, we compute how many unique apps participants choose to complete the same task. We group the search tasks by their category label (labeled by three expert annotators) and plot unique app distribution in Figure~\ref{fig:task_app_count}. Second, we analyze  Figure~\ref{fig:personal_work_count}, in which we depict how users' preference for smartphone use impacts their behavior. In this figure, we group users by their preferred use of smartphones, i.e., \textit{personal} or \textit{work} reasons. 

\partitle{Significance Testing.}
To determine significant differences we conduct the one-sided ANOVA ($p < 0.001$) test.





\section{Results and Analysis}
Here we discuss the results of the experiments described in Section \ref{sub:expSetUp}, along with other factors that can offer context to our analysis. Unless otherwise noted, reported results are significant.

\begin{figure}[h]
\vspace{-2mm}
\centering
\includegraphics[width=0.60\columnwidth]{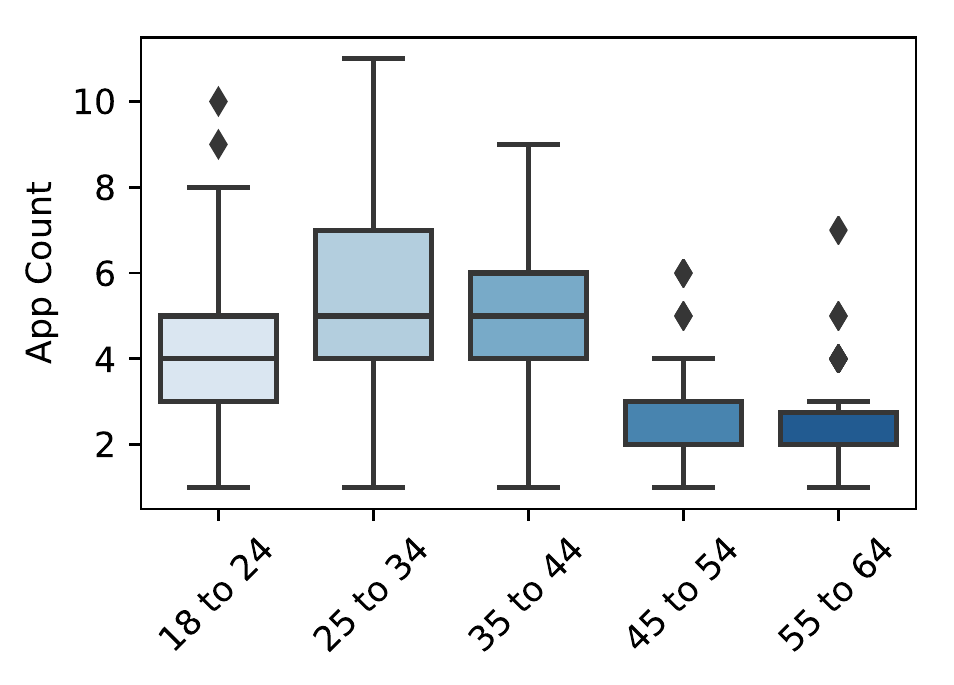}
\vspace{-2mm}
\caption{Number of apps used to complete search tasks across different age groups.}
\vspace{-5mm}
\label{fig:age_count}
\end{figure}

\partitle{Age.}
As captured in Figure \ref{fig:age_count}, younger users (18 to 44) tend to use a broader selection of apps when searching for information. This was anticipated, as older adults (45 or older) are known to be less prone to downloading and using new apps \cite{rosales2019smartphone}. Further, our results align with those reported by Gordon et al.~\cite{gordon2019app}, regarding older adults using fewer apps. Overall, the distribution of selected apps across different age groups serves as evidence to validate our hypothesis, as indeed users in varying age groups turn to a different range of apps to complete their daily search tasks.

\begin{figure}[h]
\vspace{-7mm}
\centering
\subfloat[Selected App Variety]{\includegraphics[height=4cm]{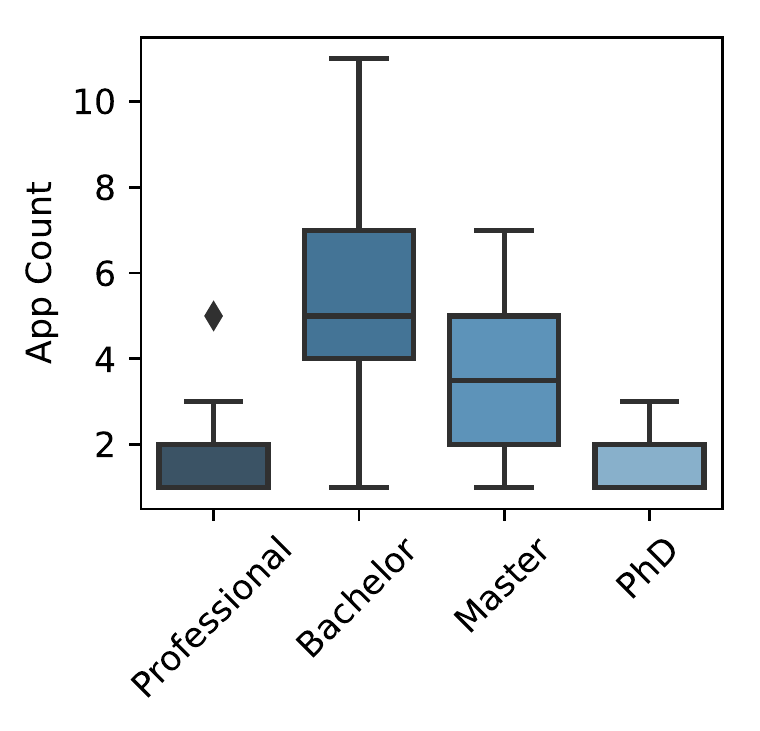}\label{fig:edu_count}}
\subfloat[App Diversity]{\includegraphics[height=4cm]{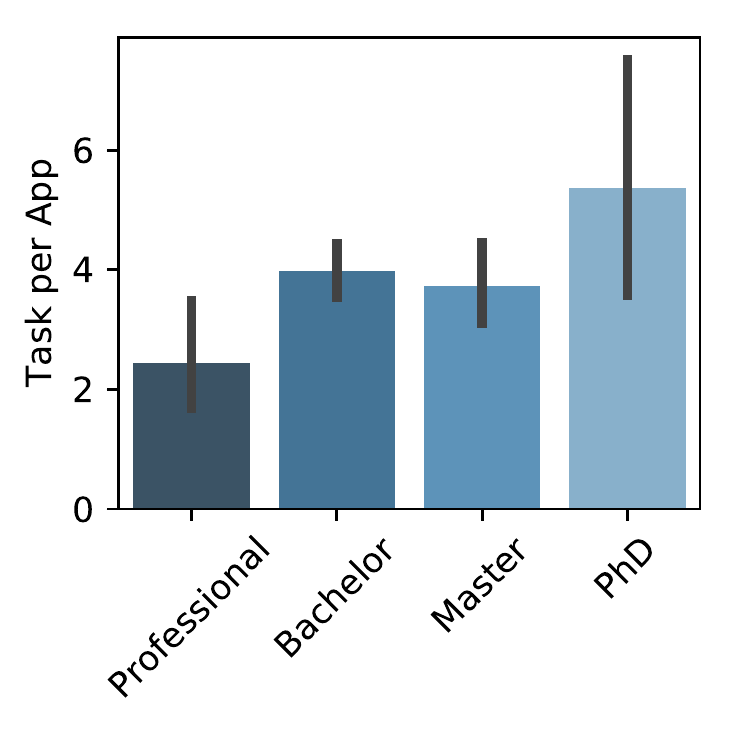}\label{fig:diversity_edu}}
\subfloat[Rare Apps]{\includegraphics[height=4cm]{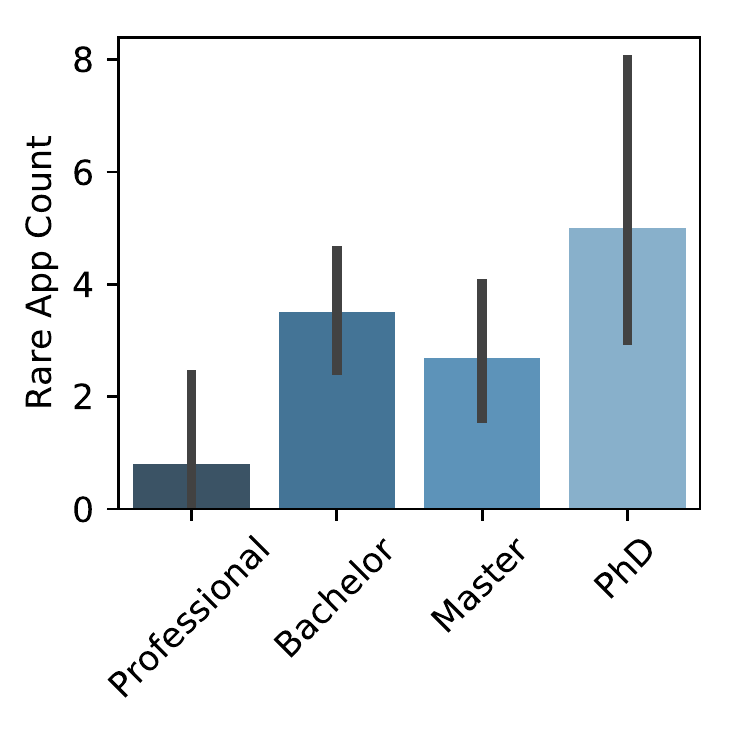}\label{fig:edu_25}}
\caption{Impact of educational background on app selection and usage.}
\vspace{-5mm}
\end{figure}

\partitle{Education.}
We posit that users' educational background could impact the manner in which they engage with mobile apps. It is evident from Figure \ref{fig:edu_count} that education background plays a role: users with a professional degree or doctorate mostly turn to a single app, a number that increases to 4 and 5 for users with a Master's degree and Bachelor's degree, respectively. This could be due to multiple reasons such as less tendency to switch between apps, or having a ``go-to'' app for all tasks. For example, as reported by Wai et al.~\cite{wai2018exploring}, bachelors are known to use a broad range of apps for learning purposes. Thus it is not unexpected to find that they choose more range of apps for searching, given their exposure to varied apps. Figure \ref{fig:diversity_edu} helps us answer this question as it shows how often users with different educational backgrounds choose a certain app, thus indicating how diverse each group is in selecting apps. We see that indeed users with a Ph.D.~degree tend to be less diverse in their selection of apps since they complete more tasks with the same app. On the other hand, we see that users with a Professional degree tend to submit fewer queries to the same app. 

We also study how much one's educational background can impact the selection of  \textit{Rare Apps}, i.e., apps that other users choose less frequently to complete the same task. We see in Figure \ref{fig:edu_25} that the rate of choosing rare apps significantly correlates with educational background. In this case, the higher the education, the more rare apps the users select. This could be due to the complexity of the tasks undertaken by users who possess higher educational backgrounds. Together, Figures \ref{fig:edu_count}, \ref{fig:diversity_edu}, and \ref{fig:edu_25} complement each other to paint a picture of how many unique choices each user group has, and how is the distribution of queries that they submit. For example, Ph.D.~and Professional users select a similar number of unique apps to complete their search tasks (Figure \ref{fig:edu_count}), however, most of the queries of Ph.D.~participants are submitted to a few apps (Figure \ref{fig:diversity_edu}) and Ph.D.~participants are very selective in the apps they choose as they have the highest rate of rare app usage (Figure \ref{fig:edu_25}).

\begin{figure}[h]
\vspace{-9mm}
\centering
\subfloat[App Count]{\includegraphics[width=0.40\columnwidth]{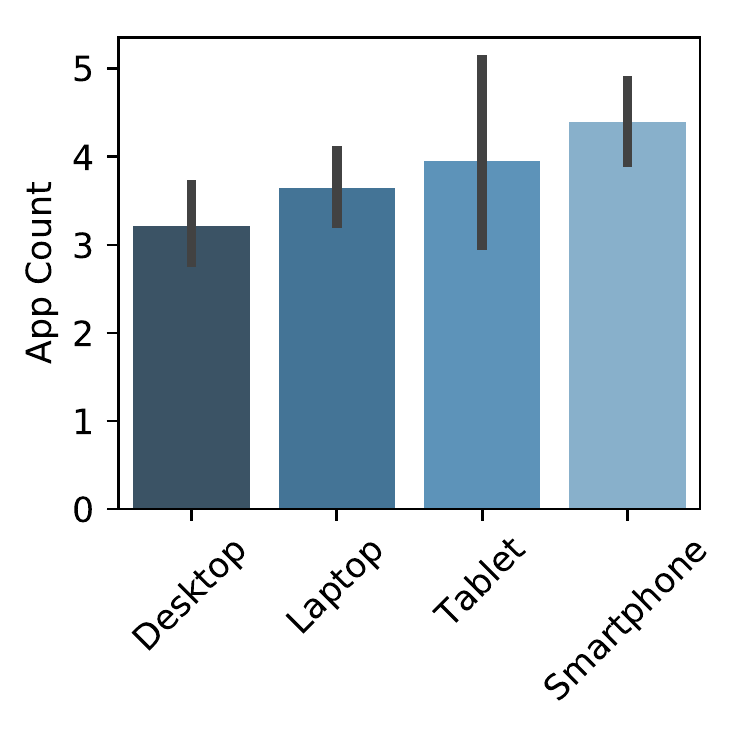}\label{fig:count_device}}
\subfloat[App diversity]{\includegraphics[width=0.40\columnwidth]{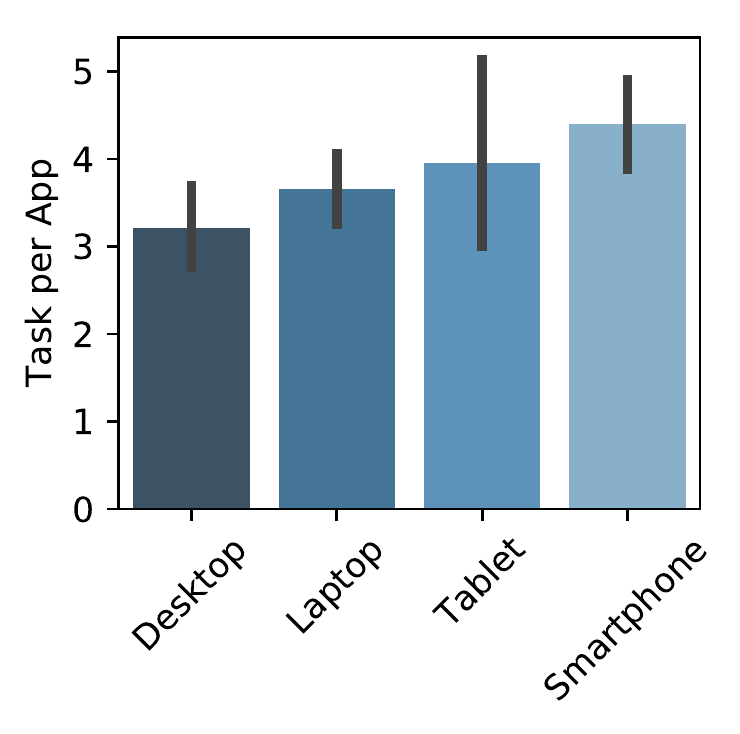}\label{fig:diversity_device}}
\vspace{-2mm}
\caption{Impact of primary device used for Internet access on selected apps.}
\vspace{-7mm}
\label{fig:device}
\end{figure}

\partitle{Device.}
We turn to Figure \ref{fig:device} for trends related to the primary device used for Internet access. It is evident from Figure \ref{fig:count_device} that the device used to access the Internet significantly impacts the number of distinct apps users turn to seek information. For example, the highest number of unique apps chosen is among smartphone users, whereas the lowest is among Desktop users. This is expected, based on recent statistics indicating that  users spend more time on smartphones than Desktops; 90\% of the time spent on smartphones is on apps \cite{saccomani_2021,broadbandsearch,timeOnlineDevice}. Moreover, we see in Figure \ref{fig:diversity_device} that smartphone users show less tendency towards diversifying their app selections (i.e., they submit more queries to fewer apps), as opposed to other users. Perhaps this is because they are more used to minimizing their search effort. Also, such users often access their smartphones in various contexts with fragmented attention~\cite{DBLP:conf/sigir/HarveyP17,DBLP:conf/chiir/AliannejadiHCPC19}, which can affect their choice of apps. 
 
\begin{figure}
\centering
\includegraphics[width=0.70\columnwidth]{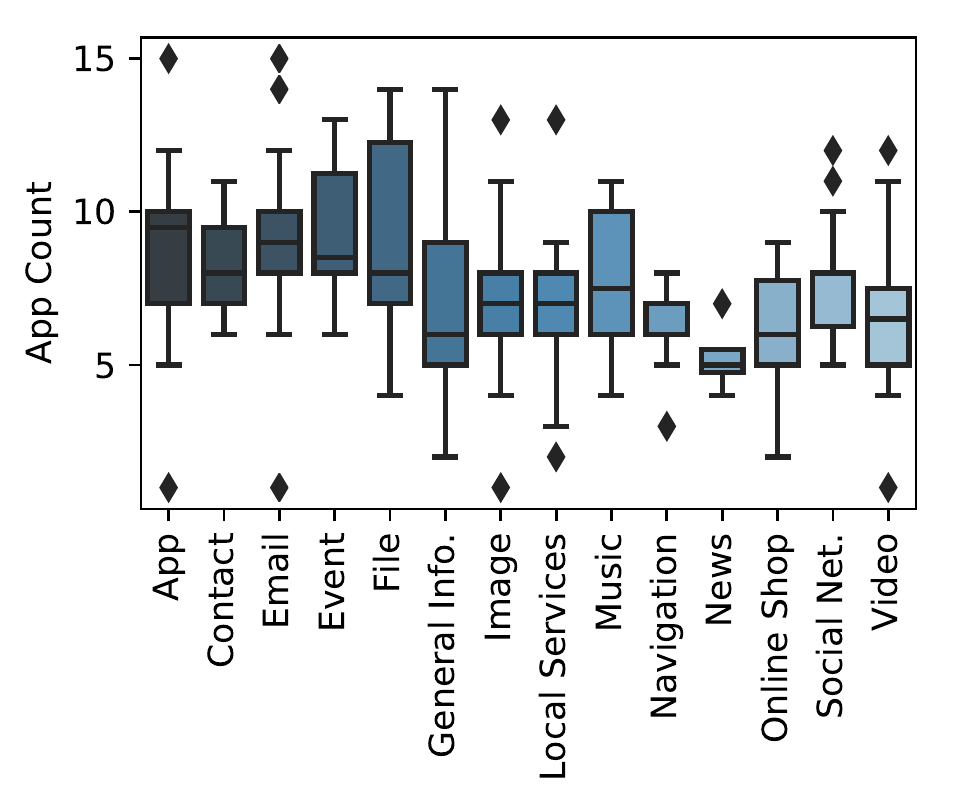}
\caption{Impact of task type on the number of apps.}
\vspace{-5mm}
\label{fig:task_app_count}
\end{figure}

\begin{figure}
\centering
\includegraphics[width=0.70\columnwidth]{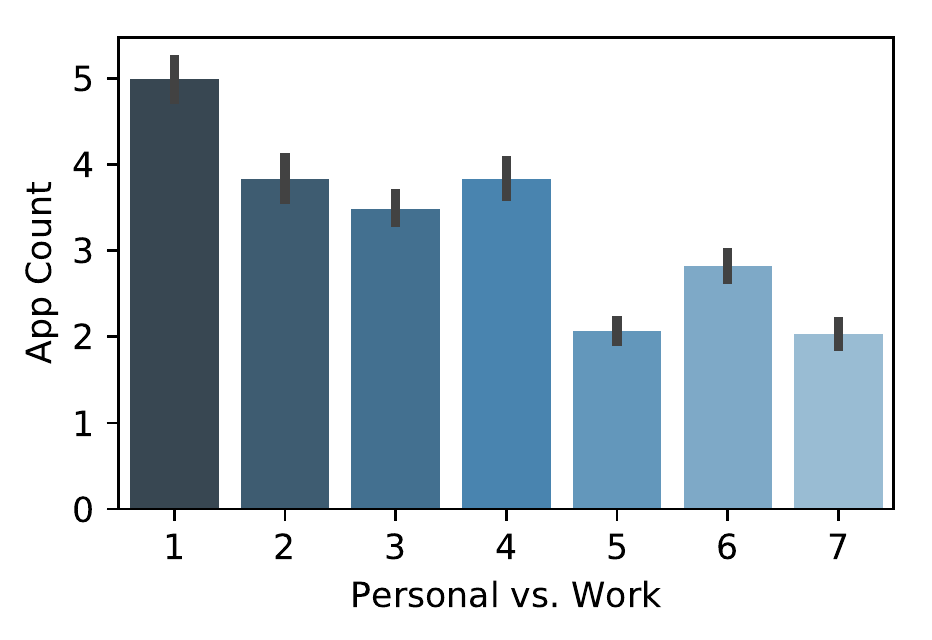}
\caption{App variety on smartphones across the personal-work spectrum. In this case, 1 indicates personal use more than work; in contrast, 7 denotes work use more than personal.}
\vspace{-2mm}
\label{fig:personal_work_count}
\end{figure}

\partitle{Task.}
We aim to understand two different aspects of the impact of task on user's behavior, i.e., personal vs.~work reasons and task type. Figure~\ref{fig:personal_work_count} shows the impact of users' search app preferences when using their smartphones mainly for personal or work purposes. We see that the more participants use their phones for personal reasons, the more apps they choose to complete their search tasks. We attribute this to the variety of tasks that they can perform for personal reasons,  which exceed those for work purposes that are generally more focused. Also, the existence of several personal apps such as instant messaging and social networking implicitly increases their choices. In order to understand this aspect better, we plot in Figure \ref{fig:task_app_count} the number of selected apps per task category. In this figure, we see that task type significantly impacts the user's behavior in terms of the apps they choose to complete the task. Interestingly we see that News and General Information categories exhibit the least number of apps, suggesting the existence of dominant apps for these categories (i.e., Google Search). However, for other task types, we observe a higher number of apps. Also, we see a larger range of app count for some task types (e.g., File), suggesting that there is a personal effect involved while completing these types of tasks, some users choosing multiple apps for these tasks while other users selecting fewer apps.

\section{Discussion and Implications}

We conclude our study by further discussing our findings and the design implications we get from them. Our discussion covers specifically how task categorization and complexity should be incorporated into the design process. Furthermore, we argue that design for users of certain age groups should be considered. This can be in the form of enhanced app switching mechanisms or app recommender systems. Finally, we point out how various contextual factors can impact user's performance and the need for further study to uncover that. 


It is of note that the current iteration of our work overlooks the two tail ends of the use spectrum, children and older adults. Both of these populations have varied levels of expertise and access when it comes to information seeking, smartphone use, and Internet access~\cite{hargittai2019internet,starkey201910}. This would suggest the need to expand our study to understand all users, not just mainstream ones. Further, it is of interest to study search tasks categories across specific use cases, for example, apps used for health information seeking~\cite{pandey2013smartphone} and learning~\cite{domingo2016exploring}, as aspects inherent to the use case could open up other areas of analysis, complementing the dimensions considered in this study, to further understand search behavior across apps. Demographic factors such as age and education clearly influence mobile app choice for search. Recommender systems could support users in their choice of the app given a specific search task, yet making available demographic profiles to these recommender systems could result in posing ``undesirable privacy risk'' \cite{wang2018toward}; further recommendation algorithms should simultaneously account for apps' permissions and users' interests and needs \cite{peng2018personalized}, which is a non-trivial task.
If we consider younger demographics, while recommender systems could indeed aid users' selection of suitable apps to turn to complete a given task, research has demonstrated that children favor knowing the source of the recommendations if they are to trust and take advantage of them \cite{pera2019little}; in their case, additional developmental traits and external factors are also a must to consider for recommendation purposes \cite{murgia2019seven}, which results in complex design requirements.
Therefore, we leave further exploration of this direction for future work.

As noted by Tian et al. \cite{tian2020identifying}, task categorization could be leveraged to improve mobile services. In our case, findings related to task categorization complemented by demographic information could provide predictive contexts to improve mobile search, allowing `` for a more personalized and engaging experience''~\cite{tian2020identifying}. 

When exploring the design space defined by our analysis, we recognize a few dimensions worth of notice: the type and complexity of the task, the effort the user is willing to put in the discovery, assessment, selection, and in case combination of suitable apps to reduce task complexity (minimum, medium and high), and the motivation for running the search (work or not). The task complexity dimension is well studied in literature and classified as an objective factor in the definition by Wildemuth et al.~\cite{wildemuth2014untangling}, who state that the task complexity is determined by the uncertain nature of the task and of the information need behind it. Also,  motivation is a well-researched dimension as it is closely linked with relevance, being a ``characteristic of all of the subjective types of relevance.''~\cite{borlund2003concept}. The app-related dimension is equally subjective and in need of further study as it accounts for a number of factors: the knowledge and degree of familiarity of the user with the available apps, the willingness to take control and engage in the selection and combination process of the apps providing the best performance, as well as the interest in finding particular apps specific to a task. That would push us to explore solutions where according to the complexity and type of the task together with the effort required to deal with it, and the motivations behind the search, users would be more or less inclined to give up control and instead trust a system to select and combine apps in order to get the best search performance.

The work reported in~\cite{tian2020identifying} goes in this  direction when they report mobile user behaviors similar to those we have discussed here in terms of engagement with multiple apps, and suggest that in the future ``intelligent switching interfaces'' would provide mobile users better search experiences. Still, we believe more study is necessary in order to better understand which apps are better suitable for which tasks and group of users. A good starting point is a work by Liu et al.~\cite{liu2016structural} who already provide some preliminary results to be expanded for aiming at a more encompassing taxonomy on demographic and task to predict user choice in the context of mobile search. While \cite{karatzoglou2012climbing} let us explore the combination of tasks and demographics into context to inform future recommending systems. 

\vspace{-2mm}

\section{Conclusions and Directions for Future Work}

Based on our analysis we can answer both RQ1 and RQ2 positively and provide some useful insights into the demographic and the extrinsic factors that condition the use of apps for search. Respectively age and education and type of device and task. Some of our findings were to be expected, e.g. younger users tend to use a broader selection of apps when searching for information and the highest number of unique apps chosen is among smartphone users. Others, such that the higher the education, the more rare apps the users select, and that the more participants use their phones for personal reasons, the more apps they choose to complete their search tasks, provide an original insight. How these factors influence each other is still to be fully studied, as for instance smartphone users often access their devices with fragmented attention~\cite{DBLP:conf/sigir/HarveyP17} and this affects the complexity of tasks they could engage with successfully. Therefore, we need to explore further the role played by task types in the selection of apps for search beyond the personal vs. work dichotomy, perhaps by focusing on task complexity instead. Starting from finding out why users do not trust apps for search to perform complex tasks and how we can change this attitude by providing users with better support when performing complex searches using smartphones. 

We need to look more closely at differences across work-related tasks as these could be linked to time pressure, fragmented attention, overall higher task complexity, and less tolerance to failure. That in turn will modify the design space making time and relevance the two dominant dimensions while pushing for more control to be left to users that could benefit from a unified mobile search and recommendation framework. Hence, in the future, we plan to conduct a field study in which users would complete mobile search tasks under various conditions and contexts, allowing us to study the impact of various contextual factors.

\bibliographystyle{splcns04}
\bibliography{References}

\end{document}